\documentclass[aps,prl,10pt,twocolumn,showpacs,superscriptaddress]{revtex4-1}
\usepackage{amsmath}
\usepackage{latexsym}
\usepackage{amssymb}
\usepackage{graphics,epstopdf}

\newcommand{\bra}[1]{\langle #1 |}
\newcommand{\ket}[1]{| #1 \rangle}

\newcommand{\tr}{\mbox{tr}}

\begin{document}

\author{Kavan Modi}
\email{kavan@quantumlah.com}
\affiliation{Centre for Quantum Technologies, National University of Singapore, Singapore}

\author{Vlatko Vedral}
\affiliation{Centre for Quantum Technologies, National University of Singapore, Singapore}
\affiliation{Department of Physics, National University of Singapore, Singapore}
\affiliation{Clarendon Laboratory, University of Oxford, Oxford UK}

\title{Unification of quantum and classical correlations and quantumness measures}
\date{\today}

\begin{abstract}
We give a pedagogical introduction to quantum discord. We the discuss the problem of separation of total correlations in a given quantum state into entanglement, dissonance, and classical correlations using the concept of relative entropy as a distance measure of correlations. This allows us to put all correlations on an equal footing. Entanglement and dissonance, whose definition is introduced here, jointly belong to what is known as quantum discord. Our methods are completely applicable for multipartite systems of arbitrary dimensions. We finally show, using relative entropy, how different notions of quantum correlations are related to each other. This gives a single theory that incorporates all correlations, quantum, classical, etc.
\end{abstract}

\maketitle

\section{Introduction}

Quantum systems are correlated in ways inaccessible to classical objects. A distinctive quantum feature of correlations is quantum entanglement \citep{PhysRev.47.777, Schrodinger:1935eq, PhysRev.48.696}. Entangled states are nonclassical as they cannot be prepared with the help of local operations and classical communication (LOCC) \cite{horodecki:865}. However, it is not the only aspect of nonclassicality of correlations due to the nature of operations allowed in the framework of LOCC. To illustrate this, one can compare a classical bit with a quantum bit; in the case of full knowledge about a classical bit, it is completely described by one of two locally distinguishable states, and the only allowed operations on the classical bit are to keep its value or flip it. To the contrary, quantum operations can prepare quantum states that are indistinguishable for a given measurement. Such operations and classical communication can lead to separable states (those which can be prepared via LOCC) which are mixtures of locally indistinguishable states. These states are nonclassical in the sense that they cannot be prepared using classical operations on classical bits.

Recent measures of nonclassical correlations are motivated by different notions of classicality and operational means to quantify nonclassicality \cite{henderson01a, PhysRevLett.88.017901, PhysRevLett.89.180402, PhysRevA.72.032317, luo:022301, PhysRevA.80.032319, PhysRevLett.104.080501}. Quantum discord has received much attention in studies involving thermodynamics and correlations \cite{zurek, PhysRevA.71.062307, Rodriguez07a, Devi-2008a, dattashaji, piani, mazzola, PhysRevLett.105.190502, PhysRevA.83.020101, cavalcantietal, pianietal}. These works are concerned with understanding the role of quantumness of correlations in a variety of systems and tasks. In some of the studies, it is also desirable to compare various notions of quantum correlations. It is well known that the different measures of quantum correlation are not identical and conceptually different. For example, the discord does not coincide with entanglement or measurement induced disturbance and a direct comparison of any two of these notions can be rather meaningless. Therefore, an unified classification of correlations is in demand as well as a unification of different notions of quantumness. In this article, using relative entropy,  we resolve some of these issues by introducing measures for classical and nonclassical correlations for quantum states under a single theory. Our theory further allows us to connect different notions of quantumness. This will allow us to generalize all measures of quantumness for  multipartite systems in symmetric and asymmetric manners. We begin with a pedagogical introduction to quantumness of correlations.

\section{Conditional entropy}

\begin{figure}[t]
\resizebox{7.67 cm}{5.87 cm}{\includegraphics{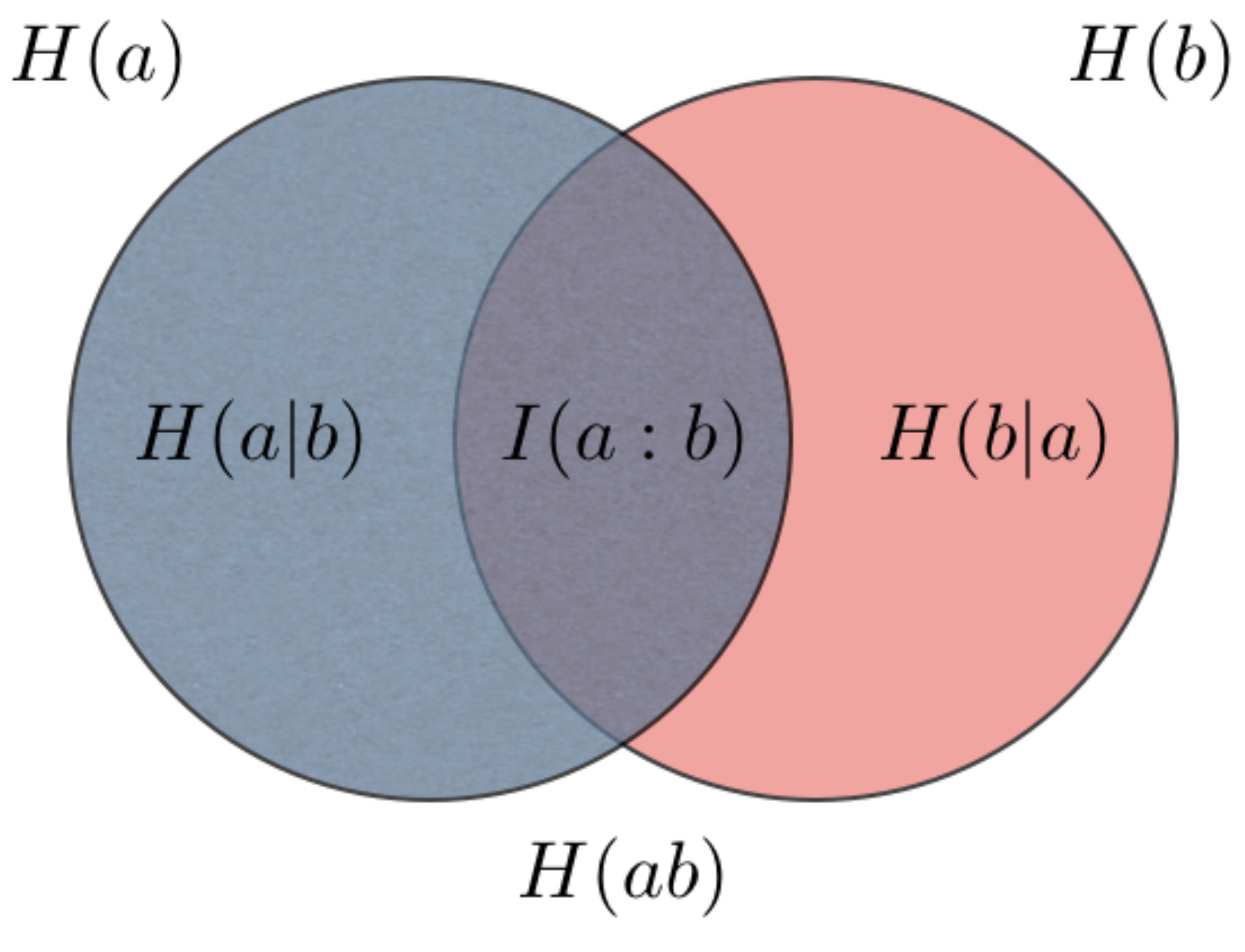}}
\caption{\emph{Conditional entropy.} The Venn diagram shows the joint entropy $H(ab)$, marginal entropies $H(a)$ and $H(b)$, conditional entropies, $H(a|b)$ and $H(b|a)$, and mutual information $I(a:b)$ a joint classical probability distribution for (correlated) random variables $a$ and $b$.\label{condent}}
\end{figure}

The story of quantumness of correlations beyond-entanglement begins with the non-uniqueness of quantum conditional entropy. In classical probability theory, conditional entropy is defined as
\begin{gather}
H(b|a)=H(ab)-H(a).\label{clce1}
\end{gather}
It is the measure ignorance of $b$ has given some knowledge of state of $a$.  Fig. \ref{condent}) depicts this relationship in a graphical manner. Another way to express the conditional entropy is as the ignorance of $b$ when the state of $a$ is known to be in the $i$th state, weighted by the probability for $i$th outcome as
\begin{gather}
H(b|a)=\sum_{i} p^a_i H(b|a=i)\label{clce2},
\end{gather}

It is the classical-equivalency of Eqs. \ref{clce1} and \ref{clce2} give rise to quantumness of correlations and in specific quantum discord \cite{datta08}. This is due to the fact that these two equations are not the same in quantum theory. While the first simply takes difference in the joint ignorance and the ignorance of $a$, the second formula depends on specific outcomes of $a$, which requires a measurement. Measurements in quantum theory are basis dependent and change the state of the system.

In generalizing the classical concepts above to quantum we replace joint classical-probability distributions with density operators and Shannon's with von Neumann's entropy. How do we deal with conditional entropy then? Clearly there are two options: Eqs. \ref{clce1} and \ref{clce2}. Let is deal with Eq. \ref{clce1} first and define quantum conditional entropy as
\begin{gather}\label{qce1}
S^{(1)}(B|A)=S(AB)-S(A).
\end{gather}
This is well known quantity in quantum information theory \cite{Schumacher} and negative of this quantity is known as coherent information. However, this is a troubling quantity as it can be negative for entangled states and for a long time there was no way to interpret the negativity. This is in start contrast with the classical conditional entropy which has a clear interpretation and is always positive.

On the other hand, we can give define the quantum version of Eq. \ref{clce2} by making measurements on party $A$. To put in the details, the joint state $\rho^{AB}$ is measured by $A$ giving $i$th outcome:
\begin{gather}
\rho^{AB}\rightarrow \sum_i \Pi^A_i \rho^{AB} \Pi^A_i = \sum_i p_i \ket{i}\bra{i} \otimes \rho^B_i,
\end{gather}
where $\Pi_i$ are rank one \emph{positive operator values measures}, $\ket{i}$ are classical flags on measuring apparatus indicating the measurement outcome, $p_i=\mbox{Tr}[\Pi^A_i \rho^{AB}]$ is probability of $i$th outcome, $\rho^B_i=\mbox{Tr}_A[\Pi^A_i\rho^{AB}]$. The conditional entropy of $B$ is then clearly defined as
\begin{gather}\label{qce2}
S^{(2)}(B|A)=S(B|\{\Pi_i\})=\sum_i p_i S(\rho^B_i),
\end{gather}
This definition of conditional entropy is always positive. The obvious problem with this definition is that the state $\rho^{AB}$ changes after the measurement. Also note that this quantity is not symmetric under party swap.

A different approach to conditional entropy is taken in \cite{PhysRevLett.79.5194, PhysRevA.60.893}, where a quantum \emph{conditional amplitude} (analogous to classical conditional probability) is defined such that it satisfies Eq. \ref{qce1}. We only mean to suggest that the two approaches above are not the only options available. Different approaches give us different distinctions of quantum theory from the classical theory. And in someway different notions of quantumness.

\section{Quantumness of correlations}

Clearly the two definition of conditional entropies above are different in quantum theory. The first one suffers from negativity and second one is `classicalization' of a quantum state. Let us now derive quantum discord and relate it to the preceding section. We start with the concept of mutual information:
\begin{gather}
I(a:b)=H(a)+H(b)-H(ab)
\end{gather}
and using Eq. \ref{clce2}
\begin{gather}
J(b|a)=H(b)-\sum_{i} p^a_i H(b|a=i).
\end{gather}
Clearly the two classical mutual information above are the same, but not in quantum theory. This is precisely what was noted by Ollivier and Zurek, and they called the difference between $I$ and $J$ \emph{quantum discord}:
\begin{align}
\delta(B|A)=I(A:B)-J(B|A).
\end{align}
Working out the details one finds that quantum discord is simply,
\begin{align}
\delta(B|A)=S^{(2)}(B|A)-S^{(1)}(B|A),
\end{align}
the difference in two definition of conditional entropy.

Henderson and Vedral \cite{henderson01a} had also looked at $J(B|A)$ called it classical correlations. In fact they advocated to that $\max_{\{\Pi_i\}}J(B|A)$ to be the classical correlations. Which meant that quantum discord is best defined as 
\begin{align}
\delta(B|A)=\min_{\{\Pi_i\}}[I(A:B)-J(B|A)].
\end{align}
Since conditional entropy in Eq. \ref{qce2} is asymmetric under party swap, quantum discord is also asymmetric under party swap.

A side note should be made at this point. The interpretation of negativity of quantum conditional entropy in Eq. \ref{qce1} was given in terms of task known as state merging~\cite{merging}, and we will see shortly that a similar task gives quantum discord an operational meaning. While the minimum of Eq. \ref{qce2} over all POVM is related to entanglement of formation between $B$ and $C$, a purification of $AB$: $E_F(BC) =\min_{\{\Pi_i\}} S^{(2)}(B|A)$~\cite{KoashiWinter}. Putting the two together leads to an task dependent operation interpretation of quantum discord \cite{cavalcantietal}. Barring the details, we can say that quantum discord between $A$ and $B$, as measured by $A$ is equal to the consumption of entanglement in state preparation of $BC$ plus state merging in those two parties. Additionally, state merging and other tasks that involve conditional entropies are asymmetric under party swap and a natural interpretation of asymmetry of quantum discord arises. 

The minimization over all POVM of quantum is not an easy problem to deal with in general. A similar quantity called \emph{measurement induced disturbance} (MID) was introduced to deal with this difficulty. MID is defined as the difference in the mutual information of a joint state, $\rho^{AB}$ and it's dephased version $\chi^{AB}$. The dephasing takes place in the marginal basis, leaving the marginal states unchanged:
\begin{align}
MID=I(\rho^{AB})-I(\chi^{AB})=S(\rho^{AB})-S(\chi^{AB}).
\end{align}
We will come back to MID later in the article.

\section{Unification of correlations}

\begin{figure}[t]
\resizebox{6 cm}{5.265 cm}{\includegraphics{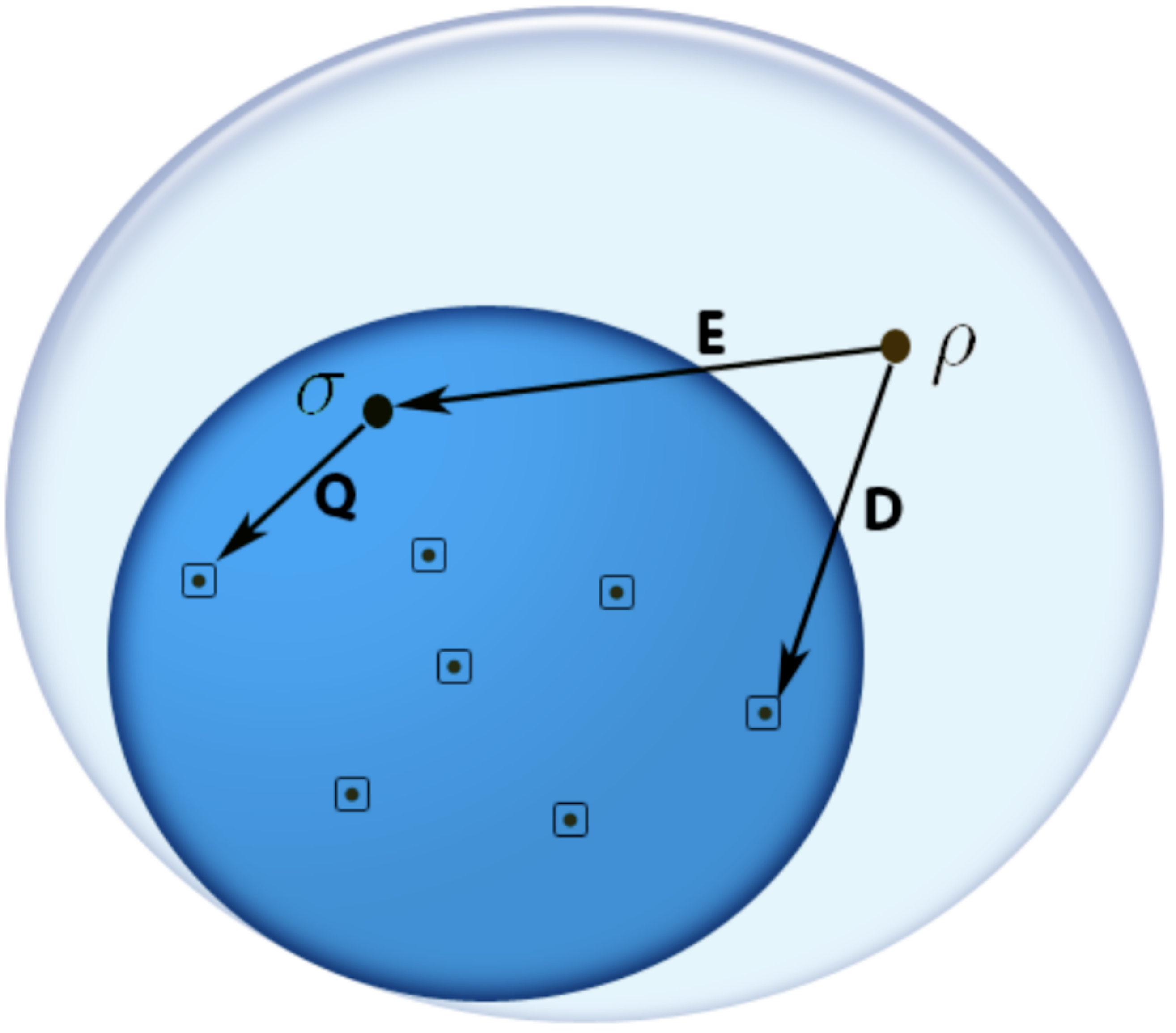}}
\caption{\emph{Correlations as a distance.}  The large ellipse represents the set of all states with the set of separable states in the smaller ellipse. The squares represent the set of classical states, and the dots within the squares are the sets of product states. $\rho$ is an entangled state and $\sigma$ is the closest separable state. The correlations are entanglement, $E$, discord, $D$, and dissonance, $Q$.\label{CORRELATIONS}}
\end{figure}

Both of the measures above are defined in terms of mutual information and therefore very difficult to generalize for multipartite case \cite{arXiv:1006.5784}. Below we will get over that hurdle by examining classical states, states that have no quantum correlations. It is easy verify that a state has zero discord and MID simultaneously. What that means is that such a state has equal value for conditional entropies in Eqs. \ref{qce1} and \ref{qce2}. Such a state is called a \emph{classical state} and has the form
\begin{gather}
\chi^{AB}=\sum_i p_i \ket{i}\bra{i}\otimes\rho^B_i
\end{gather}
when measurements are made by $A$ and $\chi^{AB}=\sum_j \rho^A_j\otimes p_j \ket{j}\bra{j}$ when measurements are made by $B$. It is then easy to see that a symmetric classical state must have the form
\begin{gather}
\chi^{AB}=\sum_{ij} p_{ij} \ket{ij}\bra{ij}.
\end{gather}
Further the conditional amplitude defined in \cite{PhysRevLett.79.5194, PhysRevA.60.893} is not a density operator and may behave strangely. In \cite{PhysRevA.81.062103} it is shown that the conditional amplitude reduces to a density operator when the state is classical.

Based on the definition of classical states we may now introduce a measure of quantum correlations as a distance from a given state to the closest classical state. The distance from a state to a state without the desired property (e.g. entanglement or discord) is a measure of that property. For example, the distance to the closest separable state is a meaningful measure of entanglement. If the distance is measured with relative entropy, the resulting measure of entanglement is the relative entropy of entanglement \cite{VPRK,VP}. Using relative entropy we define measures of nonclassical correlations as a distance to the closest classical states \cite{PhysRevLett.104.080501}, though many other distance measures can serve just as well \cite{PhysRevLett.105.190502}. We call our measure of quantum correlations \emph{relative entropy of discord}.

Since all the distances are measured with relative entropy, this provides a consistent way to compare different correlations, such as entanglement, discord, classical correlations, and \emph{quantum dissonance}, a new quantum correlation that may be present in separable states. Dissonance is a similar notion to discord, but it excludes entanglement. Lastly, here we have to make no mention of whether we want a symmetric discord measure or asymmetric, or the number of parties to be involved. We simply have to choose the appropriate classical state and no ambiguity is left. A graphical illustration is given in Fig. \ref{CORRELATIONS}.

Let us briefly define the types states discussed below. A product state of $N$-partite system, a state with no correlations of any kind, has the form of $\pi=\pi_1 \otimes \dots \otimes \pi_N$, where $\pi_n$ is the reduced state of the $n$th subsystem. The set of product states, $\mathcal{P}$, is not a convex set in the sense a mixture of product states may not be another product state. The set of classical states, $\mathcal{C}$, contains mixtures of locally distinguishable states $\chi = \sum_{k_n} p_{k_1 \dots k_N} \ket{k_1\dots k_N}\bra{k_1 \dots k_N} = \sum_{\vec{k}} p_{\vec{k}} \ket{\vec{k}}\bra{\vec{k}}$, where $p_{\vec k} $ is a joint probability distribution and local states $\ket{k_n}$ span an orthonormal basis. The correlations of these states are identified as classical correlations \cite{henderson01a, PhysRevLett.88.017901, PhysRevLett.89.180402, PhysRevLett.104.080501}. Note that $\mathcal{C}$ is not a convex set; mixing two classical states written in different bases can give rise to a nonclassical state. The set of separable states, $\mathcal{S}$, is convex and contains mixtures of the form $\sigma = \sum_{i} p_i \pi_1^{(i)} \otimes \dots \otimes \pi_N^{(i)}$. These states can be prepared using only local quantum operations and classical communication \cite{PhysRevA.40.4277} and can possess nonclassical features \cite{henderson01a, PhysRevLett.88.017901}. The set of product states is a subset of the set of classical states which in turn is a subset of the set of separable states. Finally, entangled states are all those which do not belong to the set of separable states.  The set of entangled states, $\mathcal{E}$, is not a convex set either.

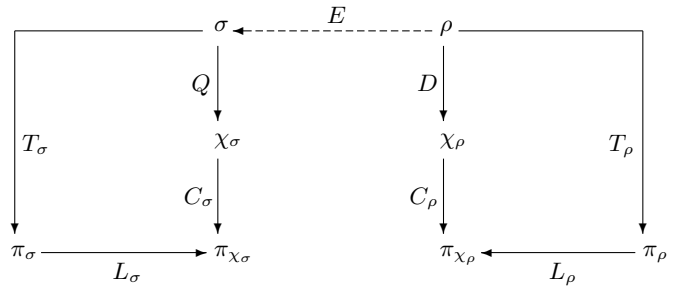
\begin{figure}[t]
	\setlength{\unitlength}{5cm}
	\begin{picture}(2.2,0.75)
	\put(0.55,0.7){$\sigma$} 
	\put(0.52,.7){\line(-1,0){.50}} 
	\put(0.02,.7){\vector(0,-1){.54}} 
	\put(0.04,0.38){$T_\sigma$}
	\put(0.56,.66){\vector(0,-1){.2}} 
	\put(.49,0.54){$Q$} 
	\put(1.15,0.7){$\rho$} 
	\multiput(1.11,.7)(-.03,0){16}{\line(1,0){.02}}
	\put(.65,.7){\vector(-1,0){.05}} 
	\put(.85,0.72){$E$} 
	\put(1.16,.66){\vector(0,-1){.2}} 
	\put(1.09,0.54){$D$} 
	\put(1.2,0.7){\line(1,0){.49}} 
	\put(1.69,.7){\vector(0,-1){.54}} 
	\put(1.6,0.38){$T_\rho$}
	\put(0.55,0.4){$\chi_\sigma$} 
	\put(0.56,.36){\vector(0,-1){.2}} 
	\put(0.47,0.24){$C_\sigma$} 
	\put(1.15,0.4){$\chi_\rho$} 
	\put(1.16,.36){\vector(0,-1){.2}} 
	\put(1.07,0.24){$C_\rho$} 
	\put(0.01,0.1){$\pi_\sigma$} 
	\put(0.09,0.11){\vector(1,0){.44}} 
	\put(0.28,0.04){$L_\sigma$} 
	\put(0.55,0.1){$\pi_{\chi_\sigma}$} 
	\put(1.15,0.1){$\pi_{\chi_\rho}$} 
	\put(1.69,0.1){$\pi_\rho$} 
	\put(1.67,0.11){\vector(-1,0){.41}} 
	\put(1.44,0.04){$L_\rho$} 
	\end{picture}
\caption{\label{ALLSTATES}\emph{Correlations in a quantum state.} An arrow from $x$ to $y$, $x \to y$, indicates that $y$ is the closest state to $x$ as measured by the relative entropy $S(x||y)$. The state $\rho\in\mathcal{E}$ (the set of entangled states), $\sigma\in\mathcal{S}$ (the set of separable states), $\chi\in\mathcal{C}$ (the set of classical states), and $\pi\in\mathcal{P}$ (the set of product states).  The distances are entanglement, $E$, quantum discord, $D$, quantum dissonance, $Q$, total mutual information, $T_\rho$ and $T_\sigma$, and classical correlations, $C_\sigma$ and $C_\rho$. All relative entropies, except for entanglement, reduce to the differences in entropies of $y$ and $x$, $S(x||y)=S(y)-S(x)$. With the aid of $L_\rho$ and $L_\sigma$ the closed path are additive.} 
\end{figure}

The relative entropy between two quantum states $x$ and $y$ is defined as $S(x||y) \equiv - \mbox{tr}(x\log y)-S(x)$, where $S(x) \equiv - \mbox{tr}(x\log x)$ is the von Neumann entropy of $x$. The relative entropy is a non-negative quantity and due to this property it often appears in the context of distance measure though technically it is not a distance, e.g. it is not symmetric. In Fig. \ref{ALLSTATES}, we present all possible types of correlations present in a quantum state $\rho$. $T_{\rho}$ is the \emph{total mutual information} of $\rho$ given by the distance to the closest product state. If $\rho$ is entangled, its entanglement is measured by the relative entropy of entanglement, $E$, which is the distance to the closest separable state $\sigma$. Having found $\sigma$, one then finds the closest classical state, $\chi_\sigma$, to it. This distance, denoted by $Q$, contains the rest of nonclassical correlations (it is similar to discord \cite{henderson01a, PhysRevLett.88.017901} but entanglement is excluded). We call this quantity \emph{quantum dissonance}. Alternatively, if we are interested in relative entropy of discord, $D$, then we find the distance between $\rho$ and closest classical state $\chi_{\rho}$. Summing up, we have the following nonclassical correlations:
\begin{align}
E =& \min_{\sigma \in \mathcal{S}} S(\rho || \sigma) 
\quad \textrm{(entanglement)}, \\\label{dis}
D =& \min_{\chi \in \mathcal{C}} S(\rho || \chi) 
\quad \textrm{(quantum discord)}, \\
Q =& \min_{\chi \in \mathcal{C}} S(\sigma || \chi) 
\quad \textrm{(quantum dissonance)}.
\end{align}
Next, we compute classical correlations as the minimal distance between a classically correlated state, $\chi$, and a product state, $\pi$: $C = \min_{\pi\in\mathcal{P}} S(\chi||\pi)$.  Finally, we compute the quantities labeled $L_\rho$ and $L_\sigma$ in Fig. \ref{ALLSTATES}, which give us additivity conditions for correlations. 

Skipping the details of the calculations (presented in \cite{PhysRevLett.104.080501}) we give the final results. The surprisingly simple results and is summarized in Fig. \ref{ALLSTATES}. First, we find that the closest classical state is obtained by making rank one POVM measurement on the quantum state
\begin{gather}
\chi=\sum_{\vec{k}} \ket{\vec{k}}\bra{\vec{k}} \rho \ket{\vec{k}}\bra{\vec{k}},
\end{gather}
where $\ket{\vec{k}}$ is projection in space at most of dimension $d^2$. To find relative entropy of discord one has to optimize over all rank one POVM. 
\begin{gather}
D= S(\rho||\chi)=\min_{\{\Pi_i\}}S(\chi)-S(\rho).
\end{gather}
Therefore finding the closest classical state is still a very difficult problem and has the same challenged as faced in computing original discord. 

We find that all correlations (except entanglement) in Fig. \ref{ALLSTATES} are given by simply taking the difference in entropy of the state at tail of the arrow from the entropy of the state at the tip of the arrow, i.e. $S(x||y)=S(y)-S(x)$ for all solid lines. Which means that a closed loop of (solid lines) yield correlations that are additive, i.e. $D+C_\rho=T_\rho+L_\rho$ or $T_\rho-C_\rho =D-L_\rho$, which is actually the original discord. We show how these two measures are related to each other below. See \cite{PhysRevLett.104.080501} for details presented in this section.

\section{Unifying quantumness measures} 

\begin{figure}[t]
	\setlength{\unitlength}{5cm}
	\begin{picture}(2.2,1.2)
	\put(.15,1.1){$\rho$} 
	\put(.16,1.06){\line(0,-1){.2}} 
	\put(.18,.94){$T$}
	\put(.13,0.8){$\pi\rho$} 
	\put(.2,1.11){\line(1,0){.37}} 
	\put(.37,1.05){$D$}
	\put(.6,1.1){$\chi_\rho$} 
	\put(.63,1.06){\line(0,-1){.2}} 
	\put(.56,.94){$C$}
	\put(.22,0.81){\line(1,0){.35}} 
	\put(.37,.83){$L$}
	\put(.6,0.8){$\pi_{\chi}$} 
	\put(.15,.66){a. Elements of }
	\put(.24,.6){quantumness }
	\put(.15,0.4){$\rho$} 
	\put(.16,.36){\vector(0,-1){.2}} 
	\put(.13,0.1){$\pi\rho=\pi_{\chi}$} 
	\put(.2,0.41){\vector(1,0){.37}} 
	\put(.6,0.4){$\chi_\rho$} 
	\put(.62,.36){\vector(-2,-1){.4}} 
	\put(.44,.22){minimize} 
	\put(.25,0){c. MID}
	\put(1.05,1.1){$\rho$} 
	\put(1.06,1.06){\vector(0,-1){.2}} 
	\put(1.03,0.8){$\pi\rho$} 
	\put(1.1,1.11){\vector(1,0){.37}} 
	\put(1.5,1.1){$\chi_\rho$} 
	\put(1.12,0.81){\vector(1,0){.35}} 
	\put(1.5,0.8){$\pi_{\chi}$} 
	\put(1.23,0.95){maximize} 
	\put(1.52,1.06){\vector(0,-1){.2}} 
	\put(1.12,.66){b. Original}
	\put(1.21,.6){discord}
	\put(1.05,0.4){$\rho$} 
	\put(1.06,.36){\vector(0,-1){.2}} 
	\put(1.03,0.1){$\pi\rho$} 
	\put(1.15,0.36){minimize} 
	\put(1.1,0.41){\vector(1,0){.37}} 
	\put(1.5,0.4){$\chi_\rho$} 
	\put(1.12,0.11){\vector(1,0){.35}} 
	\put(1.5,0.1){$\pi_{\chi}$} 
	\put(1.52,.36){\vector(0,-1){.2}} 
	\put(1.12,.0){d. RED}
	\end{picture}
\caption{\label{ALLSTATES2}\emph{Measures of quantumness.} a. Fundamental elements needed in defining a measure of quantumness of correlations. b. Original quantum discord. c. Measurement induced disturbance. d. Relative entropy of discord.}
\end{figure}
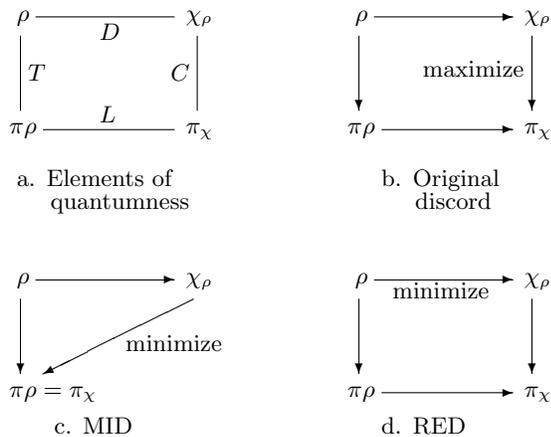

Next we note, there are four fundamental elements involved in study of quantumness of correlations. The first of it is the quantum state, $\rho$. Given a quantum state we immediately have its marginals, $\pi_\rho$. The third element is the classical state $\chi$, obtained by dephasing $\rho$ in some basis. And the final element is the marginals of $\chi_\rho$. It then turns out that different measures of quantumness are different because they put different constrain in the relationships these four elements have with each other. We have illustrated this in Fig. \ref{ALLSTATES2}.

In Fig. \ref{ALLSTATES2}a the four fundamental elements are shown. Figs. \ref{ALLSTATES2}b-d show how three measures of quantumness are found using the four elements. The original discord maximizes the distance from the classical state and its marginals. This has the meaning that the classical state is least confusing from its marginals. Quantum discord is the defined as the difference in confusion of a quantum state with its marginals and the a classical state obtained from that quantum state and it confusion with its marginals. Similarly MID attempts to minimize the confusion between the classical state and marginals of the original quantum state. This has the effect that the marginals of the quantum state are the same as the marginals of the classical state. Finally relative entropy of discord is defined as the distance between a quantum state and its closest classical state.

We now turn our attention to show that using relative entropy we can describe (and generalize) other quantumness measures such as original \emph{quantum discord} \cite{PhysRevLett.88.017901}, \emph{symmetric quantum discord} \cite{PhysRevA.80.032319}, and \emph{measurement induced disturbance}. \cite{luo:022301}. 

Vedral et al. \cite{PhysRevA.56.4452} show that quantum relative entropy has the operational meaning of being able to confuse two quantum states. The argument goes as the following: suppose you are given either $\rho$ or $\sigma$ and you have to determine which by making $N$ measurements (POVM). The probability of confusing the two states is
\begin{gather}
P_N=e^{-N S(\rho||\sigma)}.
\end{gather}

Now suppose $\rho$ is an entangled state. Then for what separable state $\sigma$ can be confused for $\rho$ the most? The answer is
\begin{gather}
P_N=e^{-N \min_{\sigma\in\mathcal{S}}S(\rho||\sigma)},
\end{gather}
where $\mathcal{S}$ is the set of separable states. This is the meaning of relative entropy of entanglement: 
\begin{gather}
	E(\rho)=\min_{\sigma\in\mathcal{S}}S(\rho||\sigma).
\end{gather}
In similar manner we can give meaning to relative entropy of discord as 
\begin{gather}
	D(\rho)=\min_{\chi\in\mathcal{C}}S(\rho||\chi)
\end{gather}
the classical state $\chi$ that imitates $\rho$ the most.

The great advantage of looking at these measures in this manner is that, now they are no longer constrained to be bipartite measures. Nor they are constrained to be symmetric or asymmetric under party exchanges. We can now define quantum discord by $n-$partite systems with measurements on $m$ subsystems. Similarly, MID can be defined in such a manner as well. The other advantage is that we know how these elements are related to each other and that there are only finite number of relationships among them that make sense, e.g. maximization of distance between a quantum state and a classical state does not make sense as on may get infinity for the result.


\section{Conclusions}

We have given a pedagogical review of ideas behind quantum correlations beyond entanglement. In doing so we were able to generalize the concepts of quantum discord to multipartite case, with no ambiguity regarding the asymmetry of quantum discord under party exchange. We have shown how three measures of quantumness can be viewed under a single formalism using relative entropy.

\begin{acknowledgements}
We acknowledge the financial support by the National Research Foundation and the Ministry of Education of Singapore. We are grateful to the organizers of the \emph{75 years of entanglement} in Kolkata for inviting our participation.
\end{acknowledgements}

\bibliography{gencorr.bib}

\end{document}